\documentclass[runningheads,a4paper]{llncs}

\usepackage{amssymb}
\usepackage{graphicx}

\usepackage{url}
\usepackage{amsmath}
\newcommand{\keywords}[1]{\par\addvspace\baselineskip
\noindent\keywordname\enspace\ignorespaces#1}

\begin{document}

\mainmatter  

\title{An arbiter PUF secured by remote random reconfigurations of an FPGA}

\titlerunning{Lecture Notes in Computer Science: An arbiter PUF secured...}

%
%
\author{Alexander Spenke$^1$,
Ralph Breithaupt$^2$\and Rainer Plaga$^2$}
\authorrunning{Lecture Notes in Computer Science: An arbiter PUF secured ...}

\institute{Hochschule Bonn-Rhein-Sieg, 53757 Sankt Augustin, Germany\\
\url{https://www.h-brs.de}
\and
Federal Office for Information Security (BSI), 53175 Bonn, Germany\\
\url{https://www.bsi.de}}

%
%

\toctitle{Lecture Notes in Computer Science}
\tocauthor{Authors' Instructions}
\maketitle

\begin{abstract}
We present a practical and highly secure method for the authentication of chips based 
on a new concept for implementing strong Physical Unclonable Function (PUF) on field programmable gate arrays (FPGA).
Its qualitatively novel feature is a remote reconfiguration in which the delay stages of
the PUF are arranged to a random pattern within a subset of the FPGA's gates. 
Before the reconfiguration is performed during authentication the PUF simply does
not exist. Hence even if an attacker has the chip under control previously she can gain no useful information about the PUF. 
This feature, together with a strict renunciation of any error correction and challenge selection criteria that depend 
on individual properties of the PUF that goes into the field make
our strong PUF construction immune to all machine learning attacks presented
in the literature. More sophisticated attacks on our strong-PUF construction will be difficult, because they require
the attacker to learn or directly measure the properties of the complete FPGA. 
A fully functional reference implementation for a secure ``chip biometrics'' is presented.
We remotely configure ten 64-stage arbiter PUFs out of 1428 lookup tables
within a time of 25 seconds and then receive one ``fingerprint'' from each PUF within 1 msec.
\end{abstract}

\section{Introduction}
"Physical unclonable functions" (PUFs) are innovative hardware devices that 
shall be hard to reproduce physically because their functionality depends on variance
in  the production or configuration process (e.g. in dopant levels)
\cite{pappu,gassend04}.
They promise to enable qualitatively novel security mechanisms e.g. for authentication                              
and key generation and distribution and have consequently become an important 
research area of hardware security\cite{puf-review,puf-review2,puf-review3}.
\\
Secure authentication of a chip when its responses
are obtained from a remote location, i.e. when its physical properties cannot be directly examined,
is an important security objective. In order to reach this objective, the chip's functionality
must be unclonable not only physically but in general (``mathematical unclonability''\cite{maes}).
This property is highly desirable e.g. for chips in banking cards and passports,
but has proven to be very difficult to ensure against well-equipped attackers on the authentication secrets
in chips\cite{tarnovsky}. 
Mathematical unclonability with PUFs can be reached with so called ``strong PUF'' which possesses a
number of challenge - response (C-R) pairs that is so large that an attacker with temporary access
to the PUF cannot evaluate them all. 
PUF constructions with an exponentially large number of 
C-R pairs have been constructed, e.g. the arbiter PUF\cite{gassend04}.
It has proved possible to construct models of such PUFs based on
a relatively small number of C-R pairs by using machine-learning programs\cite{ruehr,ruehr2,tobisch2015}.
With such a model, a simple piece of software can emulate the remote PUF,
thus breaking its security, completely.
It is the major aim of our contribution to present a qualitatively novel solution to this 
fundamental vulnerability of strong PUFs. The origin of the problem
is that the true information stored in arbiter PUFs is not
exponentially large but relatively small. 
The attacker only has to determine the relative delays of all stages in order to build a complete model.
If we estimate that the delay in one stage can be quantified by 1 byte even an XOR PUF with 10 arbiter
PUFs and 128 stages each has a true information content only about 1.3 kbyte.
It is true that this information is harder to extract than information
stored in a conventional unsecured memory. But because it is a straightforward
exercise to construct simple models in which this information appears 
as parameters it proves to be too easy to extract it. Hence we need
to require a qualitatively more difficult extraction methodology and
to increase the amount of stored information in the form
of manufacturing variations scalable and by a large factor.
\\
The basic idea to meet this requirement
is to employ a ``second challenge'' which 
specifies how the PUF is to be reconstructed with a subset of gates of an FPGA chip.
If the power of this subset is large enough, there is an super-exponentially large
number of possible PUF constructions, whose properties the attacker cannot all learn. Even if the
attacker is in physical possession of the chip on which the PUF will be
realized, she thus remains deprived of the possibility
to examine the PUF which is finally used for authentication.
\\
The security mechanism we employ for authentication is to compare 
a string of single bit responses from a PUF, 
its ``fingerprint'', with a previously recorded one from the same PUF.
We prefer this ``chip biometrics'' to authentication methods based on secret keys,
because it does not require to store any helper data for error correction 
on the chip or to select challenges based on properties derived from the chip.
These practices reveal information about properties of the PUF. 
Such information has been shown to allow very effective learning attacks on the
PUF employed in the authentication\cite{helperdata}. Because our security mechanism is
to deprive the attacker of any chance to learn anything about the authenticated PUF, it
reaches its full security potential.\\
Reconfigurable PUFs have been proposed before. Katzenbeisser et al.\cite{katzenbeisser}
and Lao and Parhi\cite{lao} studied architectures in which the challenge - response 
behaviour is changed without modifying the PUF itself. Lao and Parhi\cite{lao}
also proposed constructions in which the underlying PUF is modified in its properties.
Zhang and Lin\cite{zhang} presented a scheme against replay attacks
in which PUFs are completely reconfigured on 16 different locations
on an FPGA. Gehrer and Sigl\cite{gehrer} reconfigured PUFs on an FPGA repeatedly
to generate keys efficiently. Majzoobi et al.\cite{majzoobi2009} suggested 
the use of a ``one time PUF'' realized as a reconfigured arbiter PUF on an FPGA
that is used for a single authentication as a measure against man in the middle attacks.
Reconfiguration was not used as a measure against machine-learning attacks before.
\\
{\it Contribution}
Our main contribution is a highly practical 
and efficient PUF based authentication system that we hope reaches a security level
that rivals the best alternative
technologies for authentication. Our contributions and insights are:
\smallskip
\begin{enumerate}
\item
we develop a qualitatively new security mechanism that
prevents in principle that an attacker with temporary direct access to 
the FPGA has access to the PUF that is later used for authentication.
We thus present a strong PUF immune to all machine learning attacks presented up
to now in the literature.
\item
we demonstrate that, contrary to widespread belief, an FPGA based arbiter PUF with delay stages
based on switched multiplexers offers a viable and simple alternative to the more complex 
constructions based on delay lines that have programmable lengths;
\item
 for the first time we employ a machine learning program as a tool for 
the quantitative characterization of properties of arbiter PUFs, rather than only for predicting
its responses;
\item 
we completely avoid all risks from attacks on helper data or specially selected subsets of 
challenges by strictly only using challenges that are random relative to the chip for which they 
are chosen and employing no error correction (i.e. we perform a true ``biometrics of the chip'').
\end{enumerate}
{\it Structure}
In Section \ref{back} we supply the necessary background information
on components of our PUF construction and methods used for the characterization
of our PUF. Section \ref{design} presents first our arbiter PUF design and then our
authentication architecture. The results of an experimental characterization of
our implementation are presented in Section \ref{results}. The discussion
in Section \ref{disc} analyses the security of our construction and Section \ref{concl}
concludes.
\section{Background}
\label{back}
\subsection{Arbiter PUFs}
\label{arb1}
An arbiter PUF\cite{gassend04,helperdata,machida} consists of a chain of N pairs of multiplexers
(with an ``upper'' and ``lower'' multiplexer)
through which pass two signals that started at the same time. 
Each multiplexer pair is controlled by one
bit of a challenge of N bits. If the challenge bit is 0
the upper (lower) signal is passed through the upper (lower) multiplexer
and if the challenge bit is 1 the upper (lower) signal is 
passed through the lower (upper) multiplexer. The response
bit is 0 (1) if the lower (upper) signal arrives first at an 
arbiter at the end of the chain.
\subsubsection{Construction of arbiter PUFs on FPGAs}
\label{arb-fpga}
The construction of arbiter PUFs faces the demand 
to balance out crossing times for the two paths averaged over
the manufacturing induced fluctuations\cite{majzoobi2010,morozov}. 
On FPGAs the detailed routing on the fabric usually has to be balanced.
Compared to PUF implementations in ASICs, where routing is done by 
fixed circuit path connections, routing in FPGAs has has much more 
influence on the path delays. Due to their flexible design, a complex 
switching matrix is used to connect the logic elements to each other. 
Hence the routing delay is mostly defined by the number of switches 
involved and much less by the process variances of the gates.
While it proved possible to roughly balance
the delay within and among the delay stages by placing them symmetrically,
the delays to the first delay stage
and from the last stage to the arbiter turn out to have
imbalances due to a different routing
that are always at least an order of magnitude larger than the one
due to manufacturing variance\cite{morozov}.
If this demand is not met, the responses
are no longer unique to the individual PUF because the routing differences
are of course the same on different chips\footnote{Below ``chip'' will be a shorthand
our FPGA and ``PUF'' for one instance of our arbiter PUF construction.} for the same PUF.
Two solutions to this timing problem have
been found. The first one is to configure the lookup tables
typically provided by FPGAs as
programmable delays lines instead of multiplexers and to tune
an individual arbiter PUF by placing delay elements only in one
of the paths so that it is perfectly balanced\cite{majzoobi2010,majzoobi2012}. The other
is to duplicate the PUF on different slices of the FPGA and
to compare the output of these PUFs with identical routing
(``double arbiter PUF'')\cite{machida}.
It seems difficult to apply these solutions to
our basic approach of an arbiter PUF whose delay stages
are placed at random positions of the FPGA fabric. The
former would require to balance each individual arbiter 
for the large number of PUFs that need to be constructed. The latter solution
is not applicable if the PUF must be distributed over a considerable
fraction of the FPGA fabric as necessary for our approach.
We therefore present another solution to the 
routing problem in Section \ref{randomarb}.
\subsubsection{Learning attacks on arbiter PUFs}
\label{learn}
The simplest topological timing model of an arbiter
PUF is the following\cite{tobisch2015}. 
The parameters $\delta_0$ and $\delta_1$ are the
differences in delay time between the multiplexers of one pair
for a challenge bit of 0 and 1 respectively.
The total delay time of in a n-stage arbiter PUF $\Delta$D$_n$
is then given as: 
\begin{equation}
 \Delta D_n = \omega^T \Phi
\end{equation}
Here $\Phi$ a vector with the challenge bits as entry and
$\omega$ is the following recursive parameter:
\begin{eqnarray}
\omega_1 = \delta_{0,1} - \delta_{1,1}
\nonumber
\\
\omega_i = \delta_{0,i-1} + \delta_{1,i-1} + \delta_{0,i} - \delta_{1,i}
\nonumber
\\
\omega_{n+1} = \delta_{0,n} - \delta_{1,n}
\label{ome}
\end{eqnarray}
Here i in $\delta_{0,i}$ stands for the i-th delay stage.
It is possible to employ programs for machine learning to estimate
the vector of $\omega$ values. The estimate is often good
enough to predict the response values of an arbiter PUF
which is then completely broken as a strong PUF because it can be emulated
with a piece of software.
We used a learning program based on logistic regression together
with the RPROP optimization (Section 3 in Tobisch \& Becker\cite{tobisch2015}), to analyse
our implementation.
Because the meaning of $\omega$ is not intuitive we 
calculated the time difference of the delay difference of the
upper and lower path for a challenge bit 0 and a challenge bit 1
in each delay stage i:
\begin{equation}
\Delta\delta_i = \delta_{0,i} - \delta_{1,i} 
\label{del-del}
\end{equation}
This set of all $\Delta\delta_i$ quantifies the functionality of
the arbiter PUF.
We obtained $\Delta\delta_i$ by setting all $\delta_{0,i}$ to 0.
Then we inferred 64 $\Delta\delta_i$ values and the value of $\delta_{0,64}$
from eq.(\ref{ome}). $\Delta\delta_i$ remains dimensionless, because the absolute values
of the delay times have no influence on the responses.

\subsection{Chip biometrics}
\label{biom}
Here we authenticate chips with a protocol that is roughly analogous
to protocols for biometric authentication, e.g. with a fingerprint.
A ``basic protocol'' was discussed and realized with several types of ASIC-based
PUFs by Maes\cite{maes}.
This protocol consists of two phases, enrolment and verification. During the 
enrolment phase the verifier records a subset of responses to randomly 
chosen challenges (analogous to a subset of biometric features chosen) for
each chip to be deployed and
stores them in a database together with an ID that identifies the chip.
During the verification a chip in the field sends its identifier to the verifier.
The verifier sends one of the stored challenges. The chip determines the response to the
challenge and sends it to the verifier. The chip is verified if this response differs
by less bits than a verification threshold t from the response stored in the database.
\\
According to Maes the main drawback of the basic protocol is that it can only 
be employed in PUFs which cannot be cloned mathematically, i.e. 
which functionality cannot be cloned in principle. Our main contribution
is such a PUF, and therefore we will present a realization of the basic
protocol in section \ref{prot}.
Rather than inventing a new nomenclature (like e.g. ``FPGA signature'') we continue to use
the term ``fingerprint'' for our authenticating characteristic, but keep the quotation
marks to emphasize that this merely expresses the conceptual similarity to biometrics.

\subsection{The Smartfusion2 chip}
We used the SmartFusion2 SoC from Microsemi Corp. for our project\cite{microsemi}.
It combines a 166 MHz ARM Cortex M3 microprocessor, a system controller for
a variety of hardware tasks and interfaces, embedded non-volatile memory (eNVM)
and an FPGA fabric on the same chip. 
Because our construction needs both a microprocessor and FPGA fabric 
this SoC is ideally suited, because the housing of these components
on the same chip eliminates many possible attack vectors among these
components.
We used SmartFusion2 M2S-FG484 SOM starter
kits from Emcraft Systems for our investigations.
The FPGA of this starter kit has 12084 ``logic units''
each of which consists of a look-up table (LUT) with four
inputs, a flip-flop and a carry signal from the neighbouring logic
element.
While most of the characterizations of our implementation was performed
in JTAG programming mode, the authentication was also tested in the so called 
``in-system'' programming mode (ISP) in which the microprocessor receives data from
an interface (e.g. Ethernet and USB) and transfers it to the system controller which then
programs the FPGA and/or the eNVM. 

\section{Design of a biometric authentication system based on remote random reconfiguration}
\label{design}

\subsection{Design of a random arbiter PUF}
\label{randomarb}
\begin{figure}
\centering
\includegraphics[height=6.2cm]{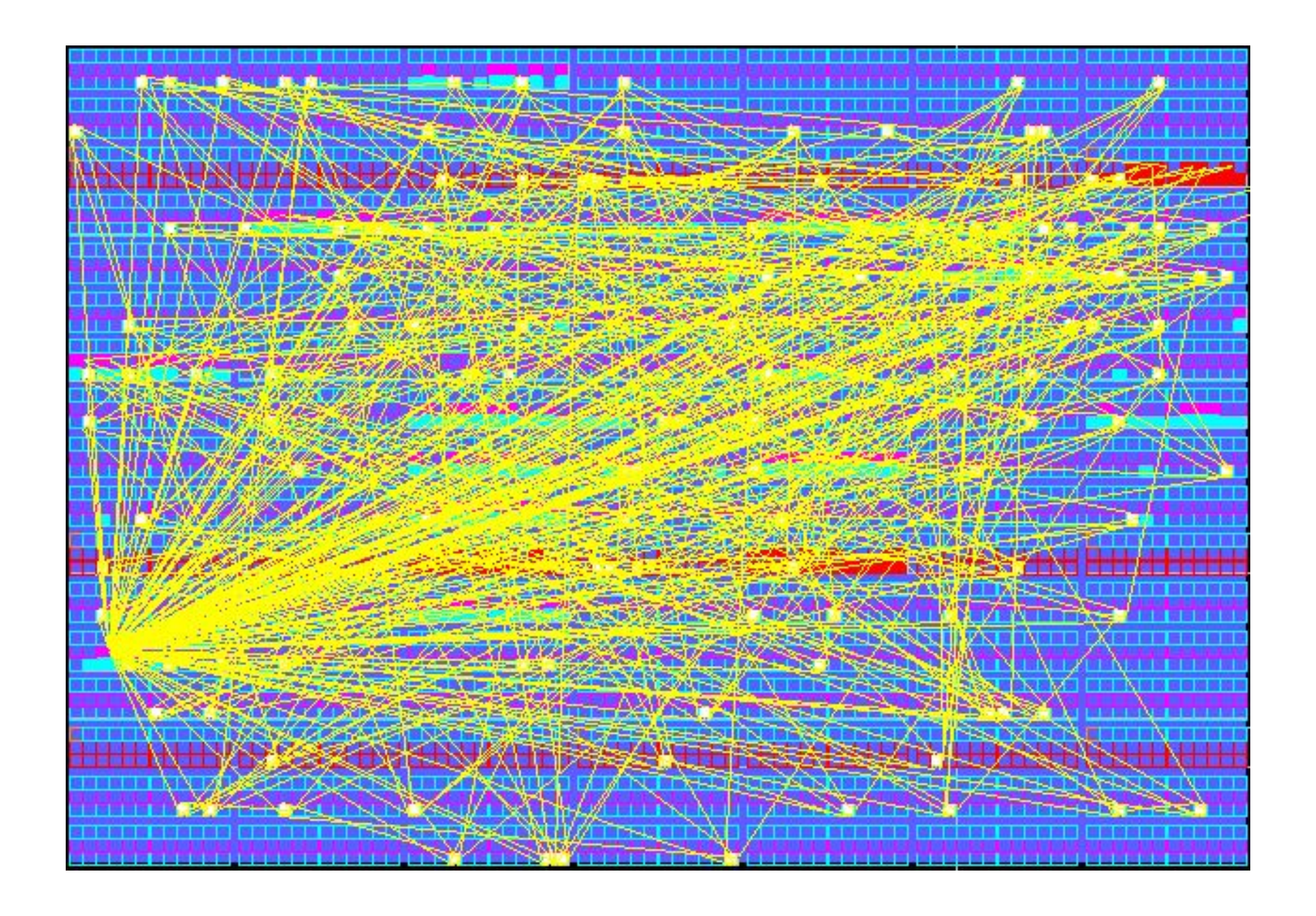}
\caption{Layout of arbiter PUF \# 1 on the region of 1428 logical units
on the FPGA. The positions of the LUTs used to implement the multiplexers
for the delay lines and the interconnections between them are displayed.}
\label{fig:plan}
\end{figure}
\begin{figure}
\centering
\includegraphics[height=6.2cm]{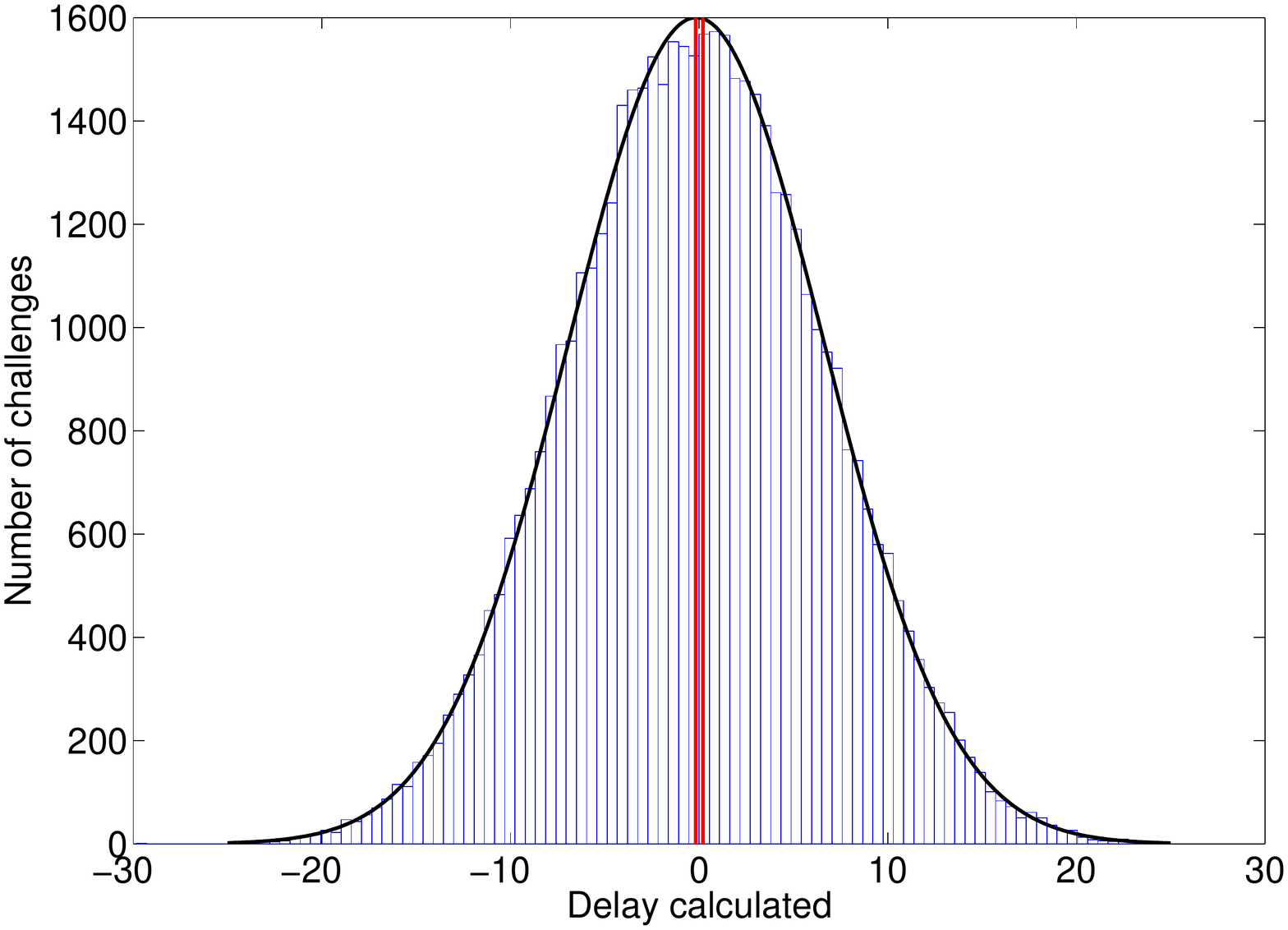}
\caption{The distribution of delay times calculated with a learning program for 50000 randomly
chosen challenges. The delay times are dimensionless because the responses do not depend on 
the absolute speed of the signals that determine them. The full curve is a Gaussian fit
to the data which has a mean value of -0.15 and a standard deviation of 6.78.
The region marked in red (light shaded)
indicates the challenges that were chosen as ``m-challenges'' because they
lead to a small delay between the paths of the arbiter PUF.}
\label{fig:del-diff}
\end{figure}

In our implementation
we realized an arbiter PUF with 64 delay stages.
We first present our solution to the problem of balanced timing 
announced in section \ref{arb-fpga}.
From a set of randomly chosen challenges we simply selected 
those challenges for which the delay-time difference between the two signals happens to be close
to 0 fortuitously. We call these challenges ``m-challenges'' (m for metastable).
We employed two methods:
\begin{enumerate}
\item
We selected challenges with metastable responses (i.e. responses that flip between
0 and 1 when the same challenge is repeatedly applied) on a ``reference chip'' that will never
leave the customer's security lab.
\\
For the m-challenges the delay difference induced by
routing and by manufacturing variance exactly balance on the reference chip. 
Therefore on other chips 
the m-challenges will also lead to delay times that are 
expected to be balanced up to time differences
induced by manufacturing variance.
\item
We modelled the reference chip
with the machine-learning model explained in section \ref{arb1}.
We then used this model to calculate the predicted delay difference
d for a given challenge. Then we selected those challenges for
which d was smaller then a maximal bound b.
\end{enumerate}
These two methods did not select the same challenges (i.e. our learning program was not precise
enough to always predict the challenges leading to metastability). When we chose b = 0.2\footnote{The upper limit has no units 
because one cannot measure the absolute delay times with machine learning programs.}
the sets selected by the two different methods had about equal power and
were both suitable for the selection of m-challenges for production.
Fig.\ref{fig:del-diff} illustrates the distribution of delay-time differences and the selection
of the bounded sample.
\\
Our construction is non-ideal because it just balances the routing delays 
(these delays will be referred to as ``routing induced delay'' below) with the delays  
due to manufacturing variance (``manufacturing induced delay'').
\\
In order to allow for a very large number of possible arbiter PUF constructions 
we selected a region of the FPGA fabric which includes of 84 $\times$ 17 = 1428 lookup tables.
We chose only a small subset of all available lookup tables to make
our scheme practical: the rest of the FPGA could still be used
for other purposes.
The 128 lookup tables used for the 64 delay stages 
of our arbiter PUF are selected randomly from this set.
The positions of the selected LUTs are stored in the ``core-cell-constraint'' file.
Fig. \ref{fig:plan} displays the layout of random PUF $\#$ 1.
\\
The decision of the response was performed in an arbiter which was not
realized as a flip-flop but with a LUT that evaluates the response R as (U AND L) OR (U AND R),
where U and L are the signal from the upper and low path of the arbiter PUF. This
construction yields a more symmetric and less temperature dependent response of the arbiter.
The VHDL code of our arbiter PUF is given in the appendix.

\subsection{Architecture and protocol of authentication system}
\label{prot}
\begin{figure}
\centering
\includegraphics[height=6.2cm]{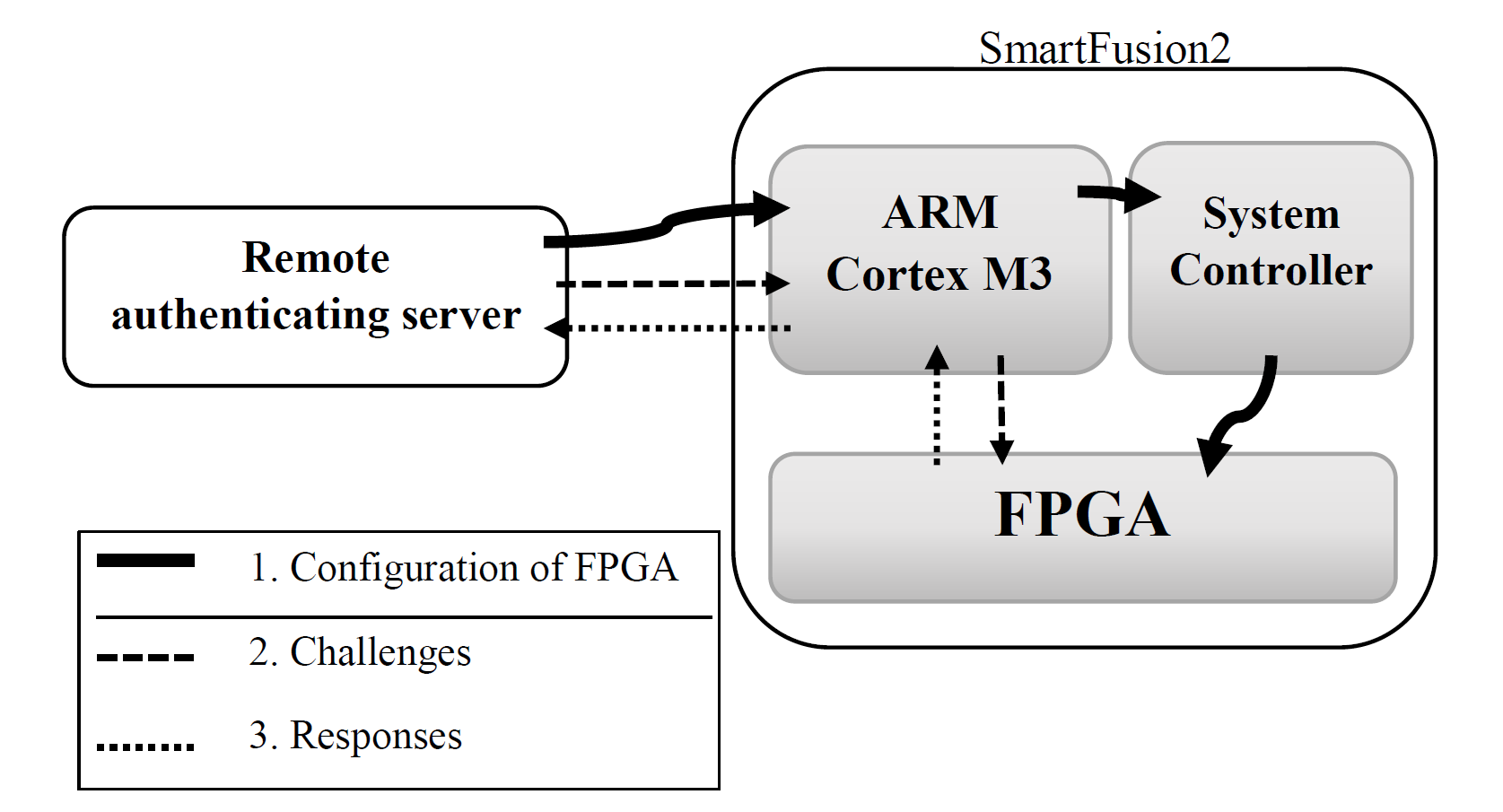}
\caption{Authentication procedure of a SmartFusion2 chip.}
\label{fig:auth}
\end{figure}

Our authentication system works analogous to conventional
biometrics and Maes' basic protocol\cite{maes} (see section \ref{biom}). 
In the enrolment phase a set of reference templates, consisting of the
responses to a number of arbiter-PUF random layouts as
``2nd challenges'', together with
100 randomly chosen m-challenges, is determined and
stored in a data base. Both these challenge-response pairs and the 
random layouts the PUFs must be kept secret.
The number of 2nd-challenge/100 m challenge pairs must be sufficiently
large for the intended application for the
chip authentication. Creating and maintaining such a database before
the deployment of the chip is a significant effort.
\\
When a chip in the field is to be authenticated, two challenges
are sent: 
\begin{enumerate}
\item
a novel type of
challenge, which consist of the compiled
VHDL code that determines the configuration of the FPGA.
This challenge, which always has a size of 556 kbyte for our FPGA\footnote{The SmartFusion2
chip does not support a partial reconfiguration of the FPGA.},
is transferred by the M3 microprocessor to the system controller
which then programs the FPGA within a time of at most 28 seconds
\footnote{With JTAG programming the total programming cycle took 25 seconds.}.
\item
100 conventional 64 bit long m-challenges that 
decide the multiplexers' settings. The 100 responses
are defined to be the ``fingerprint'' of the chip and are sent
to the authenticating party. It took about 10 $\mu$secs to obtain
a single response to an m-challenge.
\end{enumerate}
This procedure is sketched in Fig. \ref{fig:auth}.
It is identical to Maes' basic protocol except 
that instead of challenge-response pairs, 2nd-challenge and m-challenge - response
pairs have to be sent.
The authenticating party calculates the  Hamming
distance between the template and the ``fingerprint''. Only if this
 Hamming distance is smaller than a certain threshold t, the chip 
is authenticated.
\\
Both the novel and the m-challenge are analogous just to 
the information on which part of the human body (e.g. which finger) 
is to be used for authentication. 
\section{Experimental Results of Tests with the Implementation}
\label{results}
\subsection{Characterization of arbiter PUFs}
\label{arb-char}
\begin{figure}
\centering
\includegraphics[height=6.2cm]{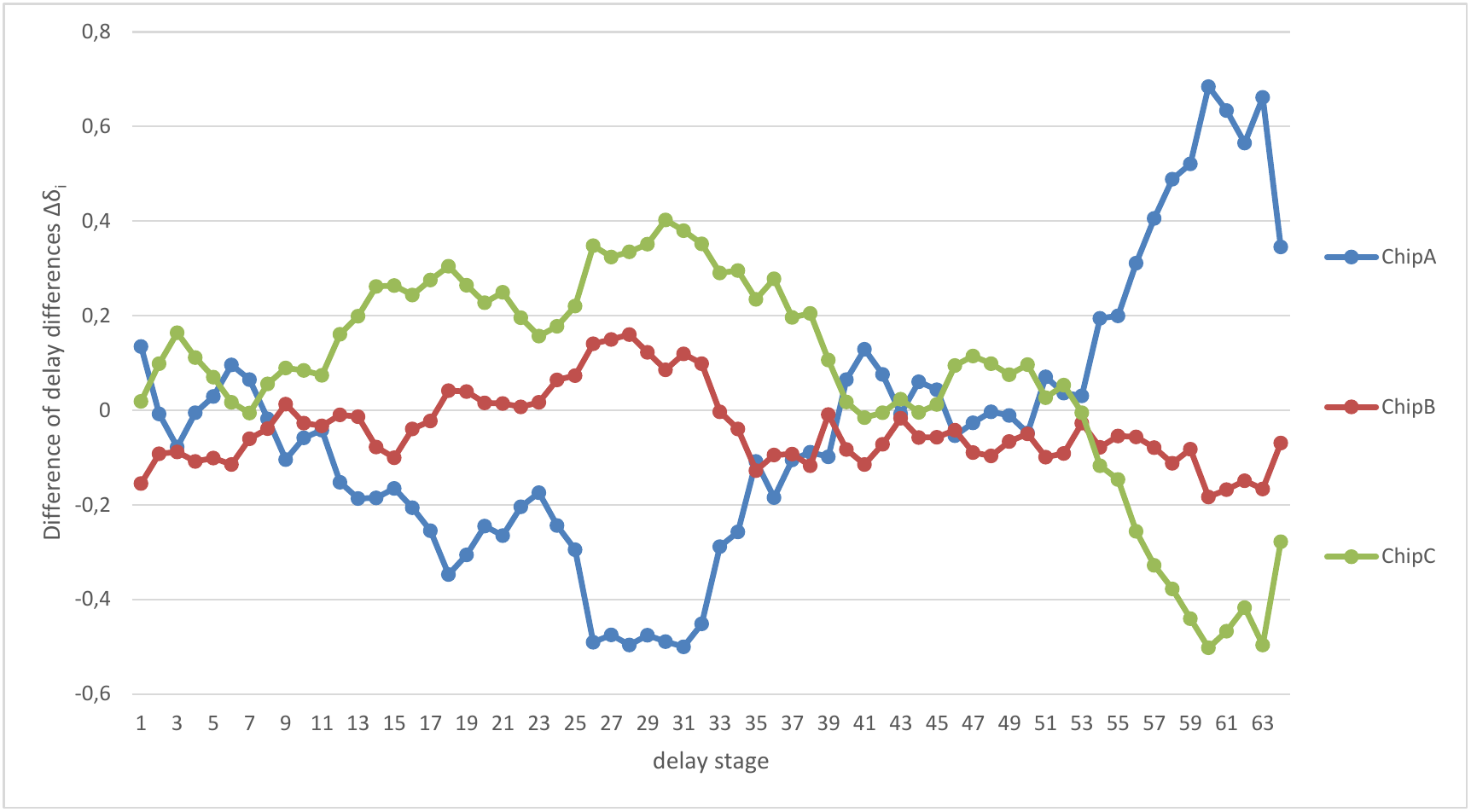}
\caption{The difference of delay differences of the 64 stages 
of one randomly placed arbiter PUF in three different chips. The delay difference are plotted
relative to the mean of the three values, i.e. only the deviation relative to the mean value is shown.}
\label{fig:deldiff2}
\end{figure}

\begin{figure}
\centering
\includegraphics[height=6.2cm]{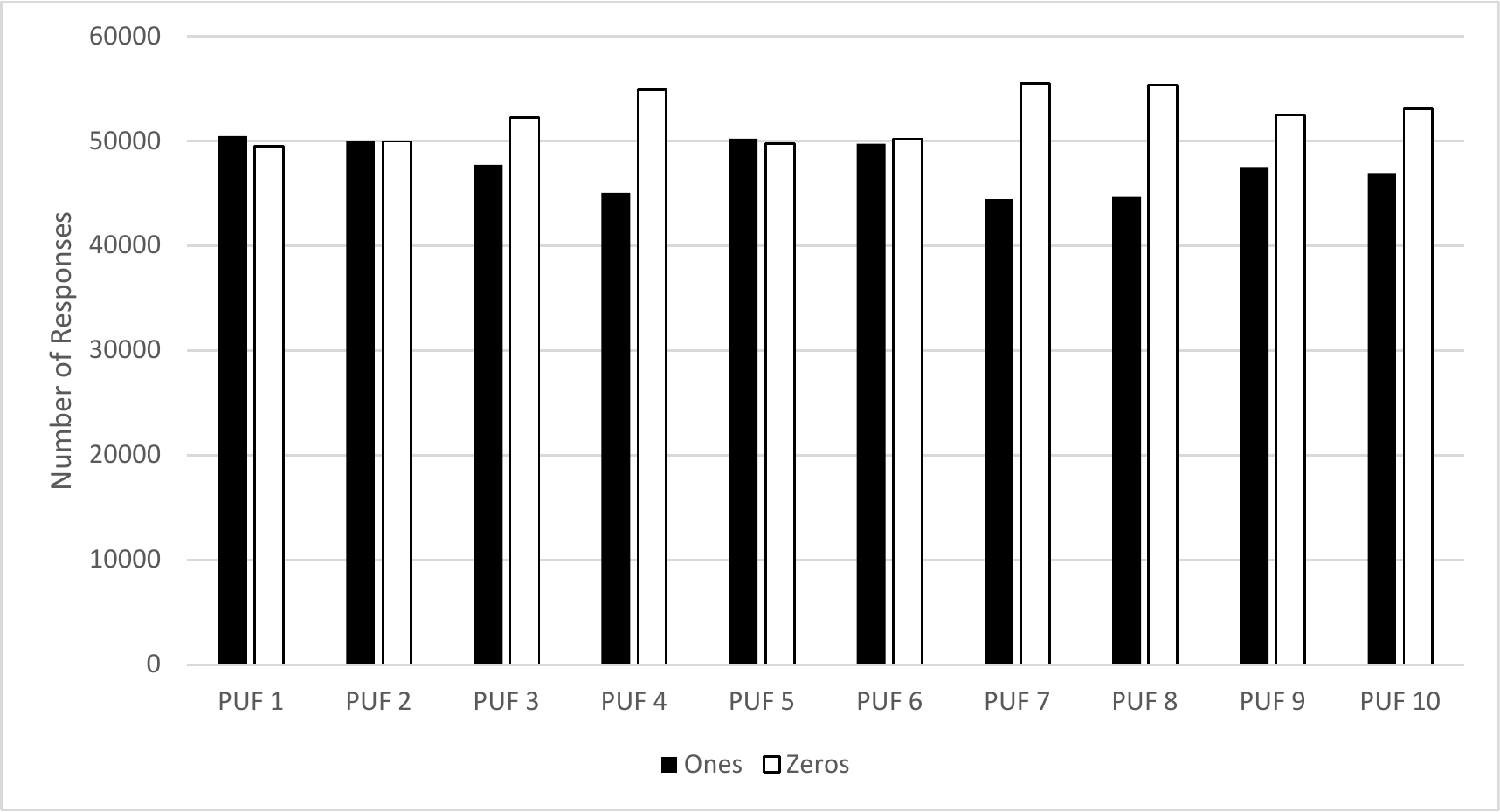}
\caption{The bias of 10 randomly placed arbiter PUFs displayed for 100000 randomly chosen
challenges.}
\label{fig:skew}
\end{figure}

We characterized the properties of ten different
randomly placed arbiter PUFs in a climate chamber at
different temperatures. 
Firstly we verified that our construction is really
a functional arbiter PUF:
\begin{enumerate}
\item
by applying the learning program discussed in Section \ref{learn} in order to test if 
our designs can be modelled as arbiter PUFs which show manufacturing
variances.
\item
by directly testing if m-challenges that lead to metastable responses
on the reference chip do mostly not lead to metastability bits in other 
chips instances due to
manufacturing variance.
\end{enumerate}
\begin{figure}
\centering
\includegraphics[height=6.2cm]{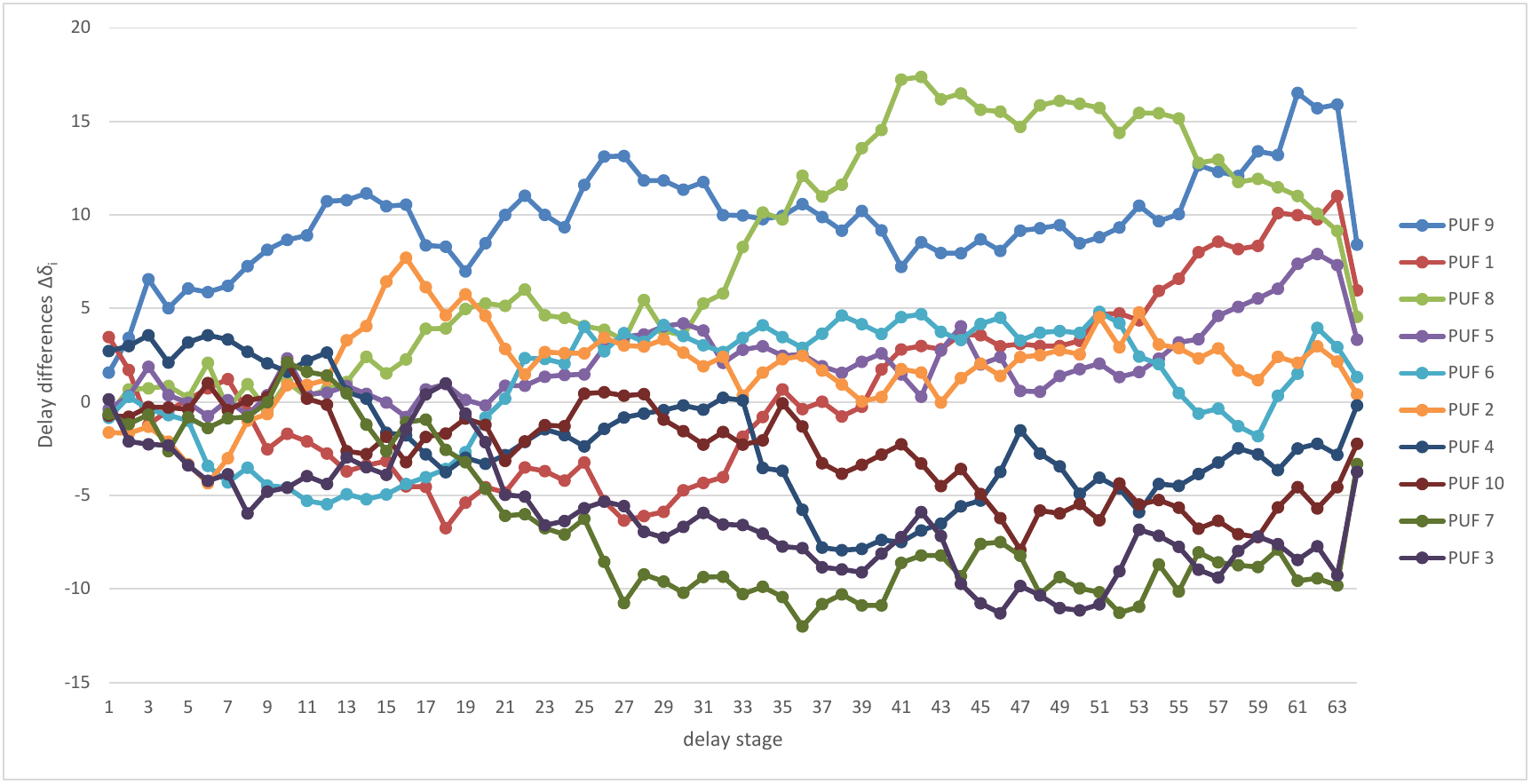}
\caption{The difference of delay differences with a challenge bit 0 and 1 of the 64 stages 
of ten randomly placed arbiter PUFs. The time is in dimensionless units
because it is derived from a machine learning program. See eq.(\ref{del-del}) for
a precise definition of the difference of delay differences.}
\label{fig:deldiff1}
\end{figure}
Fig.\ref{fig:deldiff1} shows the difference of delay differences of the 64 stages 
of ten arbiter PUFs obtained with about 20 - 30 iterations of their machine-learning program.
One recognizes that, as expected, the difference of
delays differences vary strongly among the PUFs because the routing
depends strongly on the random positions of the delay stages
on the FPGA fabric. We succeeded to predict the responses to
random challenges with an error rate of about 1.4 $\%$. 
Fig.\ref{fig:deldiff2} shows the difference of 
delay differences (see eq.(\ref{del-del}))
of the 64 stages 
of one randomly placed arbiter PUF in three different chips, relative
to the mean of the delay differences. Even though we are sure
that the derived delay differences are correct, because they enable
a correct prediction of responses, we did not achieve a deeper understanding
of their distribution, e.g. of the surprisingly strong correlation of 
the delay values in consecutive stages\footnote{We will argue below (Section \ref{disc}) that
the difficulty of understanding the routing enhances the security of our design
by obfuscation.}.
The inter-chip differences in Fig.\ref{fig:deldiff2}
are mainly due to manufacturing variance. Their mean absolute values
were found to be a factor of 29.6 smaller
than the differences among chips with a different layout
in Fig.\ref{fig:deldiff1}. This confirms the well known fact
that in a multiplexer based arbiter PUF design
the delays are dominated by differences
in the routing (Morozov et al.\cite{morozov} 
found that they dominate by a factor of 25.6 in their FPGA.)
\\
Table \ref{tab:metastables} shows the fractions of ones for 10 randomly
chosen m-challenges on two further chips. An analysis of 1000 m-challenges found
that only about 10 $\%$ of all m-challenges on chip A also lead to metastable bits on
chip B and C. Here a metastable bit is defined as a bit that flips at least
once when the challenge is applied 100000 times.
This confirms that the responses of m-challenges are strongly influenced
by manufacturing variance. Moreover this fraction is much larger 
than the one for randomly chosen challenges which we found to be 0.72 $\%$\footnote{Therefore our PUF construction
has 0.0072 $\times$ 2$^{64}$ = 1.3 $\times$ 10$^{17}$ m-challenges.}.
\begin{table}
\centering
\caption{The fraction of ones for 10 m-challenges that lead to a metastable response on chip A.
Due to manufacturing variance the r-responses mostly do not lead to metastable responses on chip B and C.
The first 10 bits of the fingerprint of chip B and C can be read from the table. If the fraction lies
between 0 and 100 \% the respective bits will be noisy.}
\includegraphics[height=4.2cm]{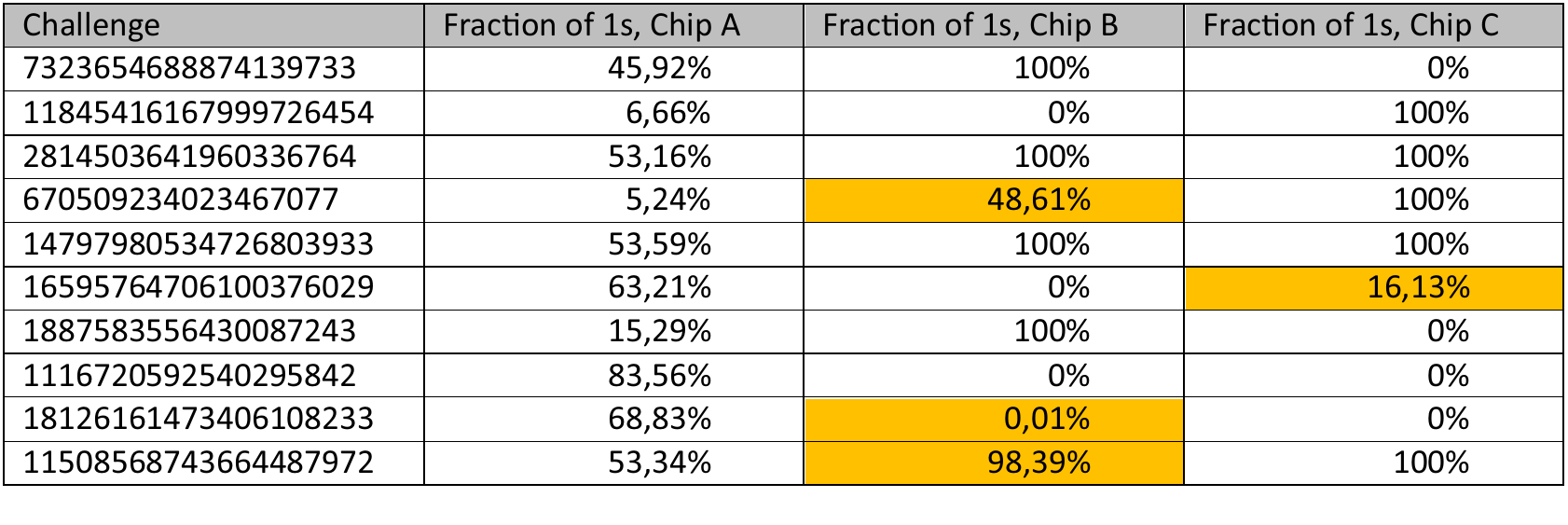}
\label{tab:metastables}
\end{table}
\\
The randomness of the responses of our PUFs was found
to depend on the placement strategy. Therefore we needed to test
uniformity, uniqueness and reliability of our PUF with
the finally chosen placement strategy that is described in Section \ref{randomarb}.
Uniformity was determined as the bias\footnote{Here we define the bias as 
${(\# {\rm \  of  \ ones}) - (\# {\rm \ of \ zeros}) \over (\# {\rm \ of \ ones}) + (\# {\rm \ of \ zeros})}$} 
of our construction displayed
(fig.\ref{fig:skew}).
The data shown in Fig.\ref{fig:skew} have
a mean bias of 4.9 $\%$, that is clearly larger than
the one expected from statistical fluctuations for our test of 0.3 $\%$
but still acceptable for fingerprints that do not have to be perfectly random. 
Moreover the bias is in a range commonly considered to be acceptable for physical random
number generators\cite{schindler}. 
\begin{figure}
\centering
\includegraphics[height=6.2cm]{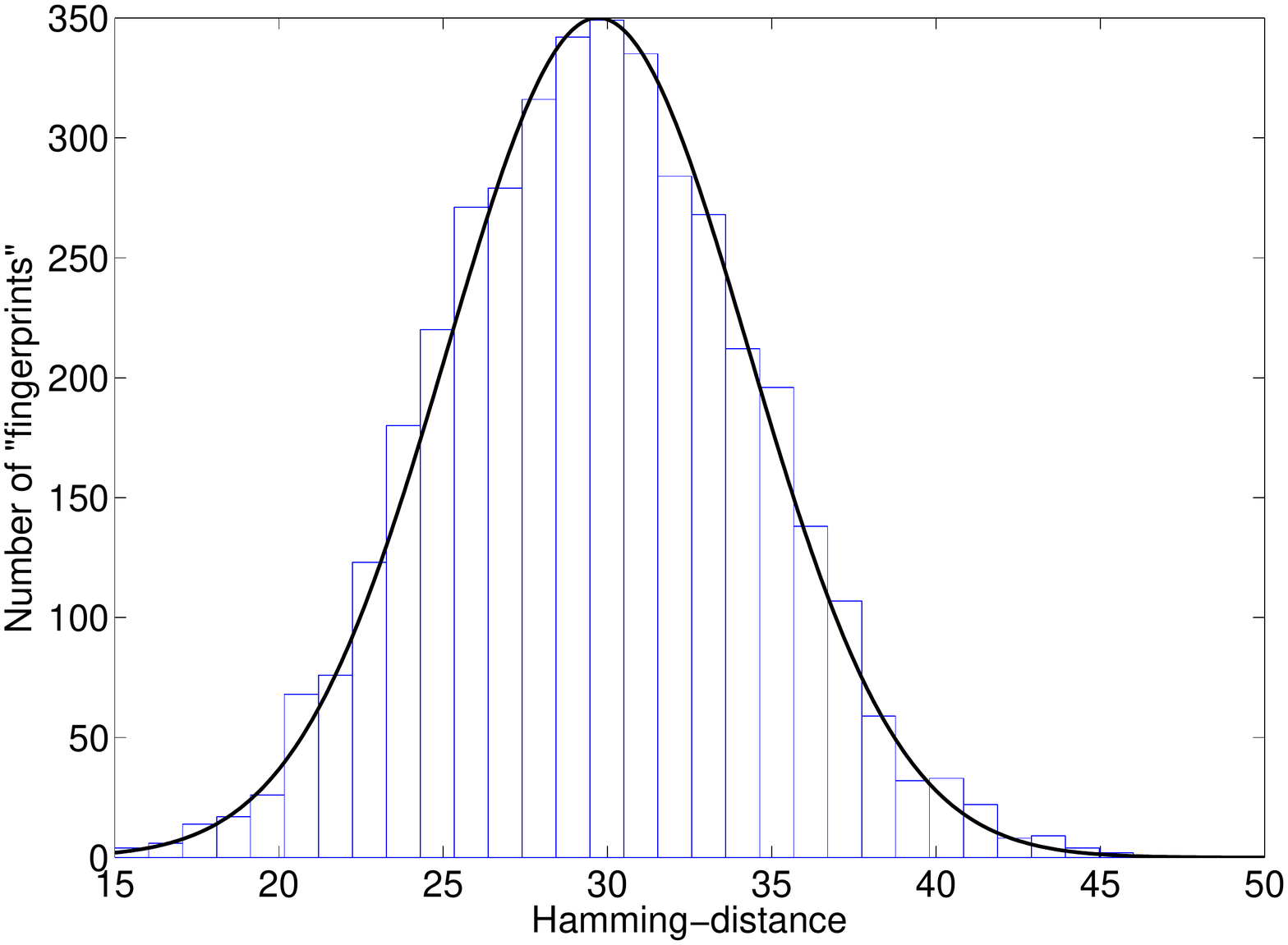}
\caption{The distribution of 4000  Hamming distances of ``fingerprint'' of chip B and C. The continuous
curve is a Gauss curve with the same mean (29.71) and standard deviation (4.57) as the data points.}
\label{fig:gdistr1}
\end{figure}
\\
The uniqueness of our PUF was quantified as
the mean  Hamming distance of a ``fingerprint'' of different chips in the same configuration
(fig.\ref{fig:gdistr1}).
It has a value of 29.7 which is significantly different from the maximal value of 50, i.e. the relative
entropy among two bits from different chips is only 0.88. This is 
not a problem for our application, as the bits in biometric templates commonly have
an entropy smaller than 1.
The reduced value can be understood as an effect of our method to choose challenges
that yield a metastable response on a reference chip. On the reference
chip (see Section \ref{randomarb}) metastability means that routing and manufacturing variation induced delay 
are exactly balanced. On the chips that are compared, the routing delay
will be the same as on the reference chip 
but the manufacturing induced delay will be different in general.
There is a 50 $\%$ chance that manufacturing induced delay between the paths
will have the same sign as the
the one of the routing induced delay on
the chips to be compared. In this case their response will always be identical.
If the delay has an opposite sign on both chips there is a 50 $\%$
chance that this will lead to a different response because
the distribution of manufacturing and routing induced delays 
in our selected sample of challenges must be the same
by design. This argument predicts a mean  Hamming distance of 25
and the value we found is similar.
The agreement of the Hamming distances induced by manufacturing variations in 
delay times in Fig.\ref{fig:gdistr1} with a Gaussian distribution is excellent.
This suggests that the bits in our ``fingerprints'' are distributed randomly, because
for the mean value of 29.7 a Gaussian is an excellent approximation to the binomial
distribution that is expected if the matching probabilities are described by a Bernoullie
process.
\\
The reliability was tested by measuring the noise in the ``fingerprint''
as a function of temperature. We found that the noise is caused exclusively by
a metastability of the arbiter that develops when
the the transit times are nearly exactly
balanced so that the both input pulses occur simultaneously. We identified
all metastable bits in a sample of 10000 challenges and its fraction
of ones $f_1$.
The probability P that metastable bit i induces a noise bit, i.e.
different responses to consecutive identical challenges is:
\begin{equation}
P_i = 2 f_i (1-f_i)
\end{equation}
The total noise fraction N determined with j metastable bits is then:
\begin{equation}
N = {{\sum_{i} P_i} \over {j}}
\end{equation}
In this manner we obtained N = 1.04 \% and 1.59 \% for two chips.
N did not change significantly with temperature in the range 5 $^o$C - 60 $^o$C.
However we found that even though its power remained roughly constant the set of metastable
bits changed with temperature because some bits became stable and others
became metastable. While the mean  Hamming distance between consecutively taken
responses with random challenges on the same PUF was 0.08 $\pm$ 0.026 $\%$
it rose to 0.35 $\pm$ 0.058 $\%$ when responses taken at 5 $^o$C and 60  $^o$C
are compared.
\subsection{FAR (interchip comparison) and FRR (intrachip comparison)}
\begin{figure}
\centering
\includegraphics[height=6.2cm]{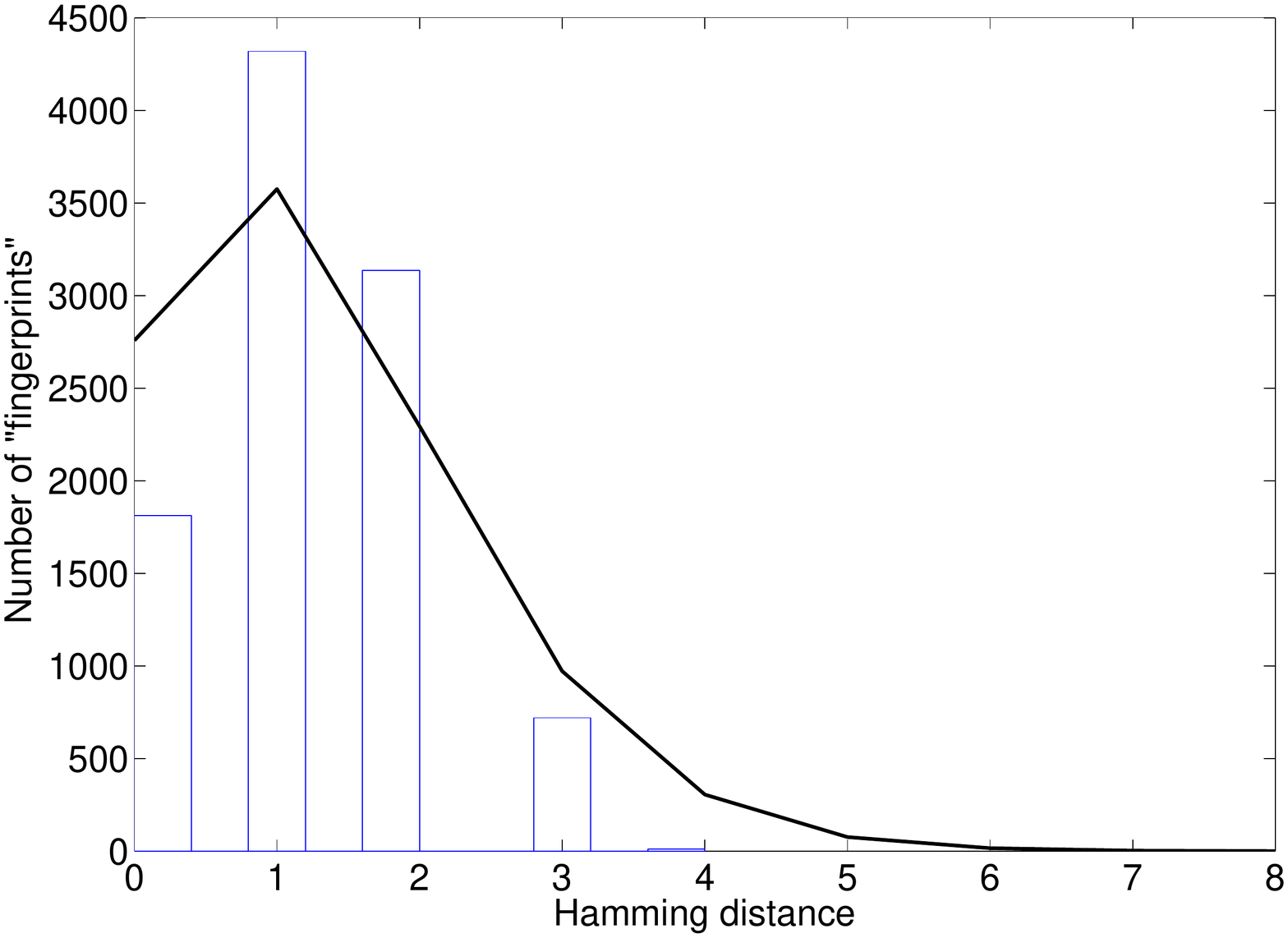}
\caption{The distribution of 10000  Hamming distances of ``fingerprint'' of chip B with each other. The continuous
curve is a fit to a binomial distribution with the same mean (1,28) as the data points.}
\label{fig:pdistr1}
\end{figure}

Analogously to the common definition in biometrics,
the false acceptance rate (FAR)
is the probability that the biometric system authenticates a
chip incorrectly and the false rejection rate (FRR) is the probability that
the system does not authenticate incorrectly.
We had seen in the previous section \ref{arb-char} that 
the distribution of matching bits in ``fingerprint'' taken from two different chips
is random and the probability for a non-match has a certain value p (p=0.297 in our case). Under these circumstances
we obtain:
\begin{equation}
FAR = \sum\limits_{i=0}^t \binom{n}{i} (1 - p)^{(n-i)} p^i
\label{binom}
\end{equation}
where t is the threshold for the number of bits 
up to which two ``fingerprints'' that
are classified a belonging to the same chip can differ. 
If we choose t = 12 we find that for our construction FAR = 2.4 $\times$ 10$^{-5}$.
The FRR is the probability that more than t bit non-matches occur
in two ``fingerprints'' of the same chip.
We estimated the FRR by determining the 10000  Hamming distances among ``fingerprints''
of the same arbiter PUF. Their distribution is plotted in Fig.\ref{fig:pdistr1}
We then performed a fit of these data to a binomial probability distribution and used
this fit to determine the FRR in a manner analogous to eq.(\ref{binom}) to FRR = 7.2 $\times$ 10$^{-9}$.
The underlying extremely conservative assumption of using a binomial distribution to fit these
data is that each bit has a mean probability of 1.3 \% to have a different value
in two consecutive measurements.  
In reality we found that the noise
for the 100 m-challenges we employed to obtain the ``fingerprint'' comes 
from six metastable bits with a fraction of ones different from 1 or 0 by more
than 0.1 $\%$. It is then much less probable to obtain a  Hamming distance
larger than 6 than expected by a binomial distribution.
As a detailed noise model is beyond the scope of the present paper
we contend ourselves with the above conservative upper bound on the FRR.
\section{Discussion of the security of our design}
\label{disc}
As a first attempt to break our construction the attacker could
try to use the 100 challenge - response pairs that were sent
to obtain the ``fingerprint'' and could be intercepted by her
to model the PUF. However we found that it took at least about
2000 challenge-response training pairs for a successful
model. It is conceivable that a smaller number might suffice
to construct a model, however it seems certain that 100 C - R pairs 
are not sufficient, because they contain an information content not larger
than 100 bits which is insufficient to encode the 64 difference of 
delay difference values that constitute the model.
\\
Another obvious attack on our construction would be 
an attempt to model all
arbiter PUFs that can be constructed when the PUF is under physical control
of the attacker. A conservative estimate of
the number of PUFs that can be constructed with our implementation
defines PUFs to be different only if they contain different gates,
i.e. all PUFs with identical gates that are only put into a different
configuration are counted as a single PUF. We then estimate the number
of PUFs N$_{PUF}$ as:
\begin{equation}
N_{PUF} = {{1428}\choose{128}} \approx 4.7 \times 10^{185}
\end{equation}
Clearly such a number of PUFs cannot even be configured on the FPGA.
Even if (theoretically) each reconfiguration could somehow 
be accelerated to take only a pico-second this would 
still take 1.6 $\times$ 10$^{166}$ years.
Therefore the only promising possibility is an attack that faithfully models 
the timing of the subset of lookup tables selected from the FPGA and the 
gates used for the routing between them. 
There are two security mechanisms that make this attack difficult.
The first one is largely due to the need for
reverse engineering: It will be more difficult
to construct a model of a complex dynamical FPGA system than of the simple static
arbiter PUF system. It seems likely that 
as a first step the attacker needs to reverse engineer the FPGA in 
order to obtain a topological model of the FPGA fabric. 
This model enables the attacker to identify
all components that influence the delays and to predict how these components
are combined in the connections between delay elements, the switching matrix
for routing and the arbiter.
Only equipped with such a construction model she will be able to understand
the distribution of the delay times of the stages we determined (but
did not understand, yet) in Section \ref{results}. 
Without such a model she would need to learn or measure the delays between
each delay element and all other delay elements, a number 
of delays that increases y with the already large
number of components.
This reverse engineering step is analogous to the one necessary
in attacks on authentication secrets stored in conventional memories and protected
by sensors or other protection mechanisms.
Once the reverse engineering is completed, 
this security mechanism is broken and further chips can be attacked with
relatively little effort.
At this point a second, PUF specific, protection mechanism kicks in:
Even on a reverse engineered FPGA the attacker needs to find out
about the manufacturing variations of the delays of all elements
of the PUF that are used in our construction.
In our implementation she needs to determine the properties
of 1428 lookup tables, i.e. the individual delays of each of them and of
all gates that are used in interconnecting them.
This makes a complete and linear characterization directly in the hardware
(e.g. with techniques developed by Tajik et al.\cite{tajik}) or
with the use of learning programs a
time-consuming task on each individual
chip that is to be modelled. This security mechanism is easily scaled: if an attacker
will succeed to break our security mechanism in an unacceptably short time, one can
increase the number of lookup tables out of which the PUFs are constructed.
In this manner our PUF construction promises to make cloning impossible
based on physical principles rather than lack of knowledge about
the protection method and technical skill to break it. 
Our second protection mechanism requires a level of effort to clone a chip
that does not significantly decrease when the protection mechanism is
fully understood by the attacker.

\section{Conclusion}
\label{concl}
We presented a qualitatively novel concept to increase the security of strong PUFs.
Up to now most attempts to make PUFs more secure aimed at making the individual
PUF construction more complex, e.g. by performing an XOR between several PUFs. 
This strategy is limited by the need to keep the final output sufficiently reliable.
Our strategy was to keep the individual PUF simple but to force the attacker to model
not only the static PUF but a part of a dynamical FPGA system. This concept enabled
a qualitative increase the complexity of the system that has to be modelled compared to
previous constructions. The only fundamental limit to increasing it further is the
available size of the FPGA fabric. 
Our FPGA-based arbiter PUF design
itself is simpler than the ones proposed up to now. 
The price
one has to pay for the gain in security is an additional overhead for the sending
of the ``2nd challenge'' that specifies a reconfiguration of the PUF.  However, it is not necessary
to introduce this overhead for each authentication. From the 1428 LUTs assigned to our
construction in our implementation
it is possible to construct 10 arbiter PUFs with one second challenge, so that 
only every 10th authentication needs the additional overhead. 

\subsubsection*{Acknowledgements.} 
We thank Georg Becker, Shahin Tajic, Jean-Pierre Seifert and Marco Winzker for helpful discussions.
Georg Becker kindly provided a copy of his machine-learning program to us.

\section*{Appendix}
\noindent
{\it VHDL Code for our arbiter PUF construction. ``above'' and ``below'' stand for the
upper and lower signal pathes. [...] stands for the insertion of 62 additional consecutive, identical sub-parts
of the code.}
\begin{verbatim}
----------------------------------------------------------------------------
----
-- Company: XXX
-- File: Arbiter_PUF.vhd
-- Description:
-- Arbiter Physical Unclonable Function (PUF)
-- Submodul to evaluate response from Arbiter PUF.
-- The input challenge defines the connection of a row of different gates.
-- An Arbiter at the end of this gates evaluates which of the two signals
arrived first
-- and sets the corresponding response.
-- Targeted device: <Family::SmartFusion2> <Die::M2S150> <Package::FG1152>
-- Author: XXX
-- Date: 12.2015
----------------------------------------------------------------------------
----
library IEEE;
use IEEE.std_logic_1164.all;
use IEEE.numeric_std.all;

entity Arbiter_PUF is
port (
c : IN std_logic_vector(63 downto 0); -- challenge
enable : IN std_logic; -- enable signal for arbiter puf
dc : IN std_logic; -- don't care input for LUTs
ready : OUT std_logic; -- ready signal
r : OUT std_logic -- response
);
end Arbiter_PUF;

architecture architecture_Arbiter_PUF of Arbiter_PUF is
-- signal, component etc. declarations
attribute syn_keep : boolean;
signal above : std_logic := '0';
signal c0 : std_logic := '0';
signal above0,above1, [...],above64 : std_logic := '0'; --
top arbiter puf signals
signal below : std_logic := '0';
signal below0,below1, [...] ,below64 : std_logic := '0'; --
bottom arbiter puf signals
-- set syn_keep for PUF signals to prevent removing in synthesis optimization
attribute syn_keep of above,above0,above1, [...] ,above64,
below,below0,below1, [...]
,below64,c0 : signal is true;
begin
-- architecture body
above0 <= above when (c0= '0' and dc = '0') else below;
below0 <= below when (c0= '0' and dc = '0') else above;
-- challenge 0
above1 <= above0 when (c(0)= '0' and dc = '0') else below0;
below1 <= below0 when (c(0)= '0' and dc = '0') else above0;
-- challenge 1
above2 <= above1 when (c(1)= '0' and dc = '0') else below1;
below2 <= below1 when (c(1)= '0' and dc = '0') else above1;
[...]
-- challenge 63
above64 <= above63 when (c(63)= '0' and dc = '0') else
below63;
below64 <= below63 when (c(63)= '0' and dc = '0') else
above63;

---- Arbiter to generate response
r <= (below64 and not(above64)) or (below64 and r);
-- ENABLE PROCESS
process--(enable)
begin
wait on enable;
if(enable = '1') then
above <= '1';
below <= '1';
-- wait until response is generated
wait on r;
ready <= '1';
else -- enable = '0'
above <= '0';
below <= '0';
ready <= '0';
end if;
end process;
end architecture_Arbiter_PUF;




\end{verbatim}
\end{document}